\documentclass[prl,superscriptaddress,floatfix,preprintnumbers,twocolumn]{revtex4}

\usepackage{url,color}
\usepackage{graphicx}
\usepackage{bm} 
\usepackage{amsfonts,amsmath,amssymb,amsbsy,latexsym}

\newcommand{\bc}{\begin{center}}
\newcommand{\ec}{\end{center}}
\newcommand{\be}{\begin{equation}}
\newcommand{\ee}{\end{equation}}
\newcommand{\ba}{\begin{eqnarray*}}
\newcommand{\ea}{\end{eqnarray*}}
\newcommand{\bna}{\begin{eqnarray}}
\newcommand{\ena}{\end{eqnarray}}

\newcommand{\mpaa}{\begin{minipage}[t]{7.5cm}}
\newcommand{\mpea}{\end{minipage}}

\begin{document}

\title{Anomalous diffusion in random dynamical systems}

\author{Yuzuru Sato}
\email{ysato@math.sci.hokudai.ac.jp}
\affiliation{RIES / Department of Mathematics, Hokkaido University, 
N20 W10 Kita-ku, Sapporo, 0010020 Hokkaido, Japan}
\affiliation{London Mathematical Laboratory, 14 Buckingham Street, London WC2N 6DF, UK}
\author{Rainer Klages}
\email{r.klages@qmul.ac.uk}
\affiliation{Queen Mary University of London, School of Mathematical Sciences, Mile End Road, London E1 4NS, UK}
\affiliation{Institut f\"ur Theoretische Physik, Technische Universit\"at Berlin, Hardenbergstra{\ss}e 36, 10623 Berlin, Germany}
\affiliation{Institute for Theoretical Physics, University of Cologne, Z\"ulpicher Stra{\ss}e 77, 50937 Cologne, Germany}
\date{\today}

\begin{abstract}

  Consider a chaotic dynamical system generating Brownian motion-like
  diffusion. Consider a second, non-chaotic system in which all
  particles localize. Let a particle experience a random combination
  of both systems by sampling between them in time.  What type of
  diffusion is exhibited by this {\em random dynamical system}? We
  show that the resulting dynamics can generate anomalous diffusion,
  where in contrast to Brownian normal diffusion the mean square
  displacement of an ensemble of particles increases nonlinearly in
  time. Randomly mixing simple deterministic walks on the line we find
  anomalous dynamics characterised by ageing, weak ergodicity
    breaking, breaking of self-averaging and infinite invariant
    densities. This result holds for general types of noise and for
    perturbing nonlinear dynamics in bifurcation scenarios.

\end{abstract}


\maketitle

Many diffusion processes in nature and society were found to behave
profoundly different from Brownian motion, which describes the
random-looking flickering of a tracer particle in a fluid
\cite{SZK93,KSZ96,KRS08,MJCB14,HoFr13,MeSo15,VLRS11,ZDK15}.  Brownian
dynamics provided a long-standing powerful paradigm to understand
spreading in terms of {\em normal diffusion}, where the mean square
displacement (MSD) of an ensemble of particles increases linearly in
the long time limit, $\langle x^2\rangle\sim t^{\alpha}$ with
$\alpha=1$. {\em Anomalous diffusion} is characterized by an exponent
$\alpha\neq1$ \cite{SZK93,KSZ96,KRS08,MJCB14}.  Subdiffusion with
$\alpha<1$ is commonly encountered in crowded environments as, e.g.,
for organelles moving in biological cells and single-file diffusion in
nanoporous material \cite{HoFr13,MeSo15}.  Superdiffusion with
$\alpha>1$ is displayed by a variety of other systems, like animals
searching for food and light propagating through disordered matter
\cite{VLRS11,ZDK15}.

Experimental data exhibiting anomalous diffusion is often modelled
successfully by advanced concepts of stochastic theory, most notably
subdiffusive continuous time random walks, superdiffusive L\'evy
walks, generalized Langevin equations, or fractional Fokker-Planck
equations \cite{SZK93,KSZ96,KRS08,HoFr13,MeSo15,VLRS11,ZDK15,MJCB14}.
In these stochastic models the mechanisms generating anomalous
diffusion are put in by hand on a coarse grained level, either via
non-Gaussian probability distributions or via power law memory
kernels. While this {\em stochastic} approach to anomalous diffusion
has matured impressively, anomalous diffusion in {\em deterministic}
dynamical systems is yet poorly understood. In nonlinear deterministic
equations of motion there are only few mechanisms known to generate
anomalous diffusion \cite{KRS08}: stickiness of orbits to KAM tori in
Hamiltonian dynamics \cite{ZGNR86,SZK93,KSZ96,Zas02}, marginally
unstable fixed points in dissipative Pomeau-Manneville-like maps
\cite{GeTo84,ZuKl93a,Bar03,KCKSG06,KKCSG06} and non-trivial topologies
exhibited by polygonal billiards \cite{Kla06}. In this Letter we
introduce a simple hybrid system at the interface between
deterministic and stochastic dynamics. We show that it yields another
generic mechanism for anomalous diffusion based on stochastic chaos in
random dynamical systems \cite{FSSWPDD17,SDLR19}. This sheds new light
on the microscopic origin of anomalous dynamics. Similar models
  have been used to understand the convection of particles in flowing
  fluids \cite{LOC90} including fractal clustering \cite{WMGW12} and
  path coalescence \cite{WiMe03}, the localisation transition in
  continuum percolation problems \cite{HFF06}, intermittency in
  nonlinear electronic circuits \cite{HPHH94} and random attractors in
  stochastic climate dynamics \cite{CSG11}. Accordingly, we expect
  fruitful applications of our approach to these problems.

\begin{figure}[htpb]
    \centering
\includegraphics[scale=0.5]{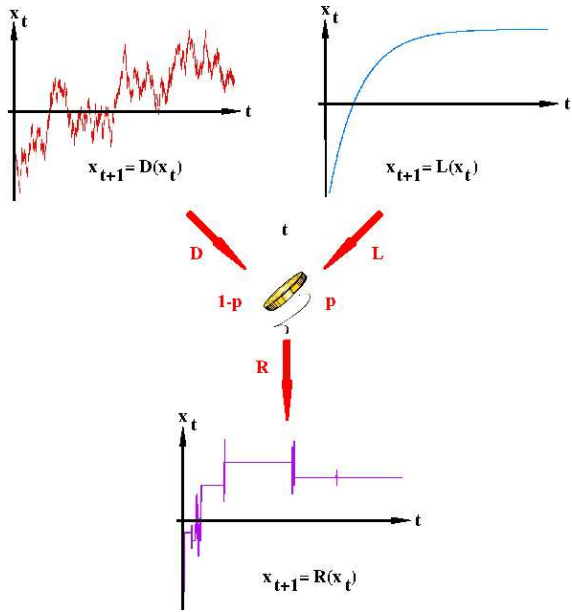}
\caption{Diffusion generated by a random dynamical system. The
  three time series in space-time plots display the position
  $x_t\in\mathbb{R}$ of a point particle at discrete time
  $t\in\mathbb{N}$.  The trajectory in the upper left is generated by
  the equation of motion $x_{t+1}=D(x_t)$ using a deterministic
  dynamical system $D$ that yields normal diffusion. The trajectory in
  the upper right is from $x_{t+1}=L(x_t)$ for a deterministic
  dynamical system $L$ where all particles localize in space. The {\em
    random dynamical system} $R$ mixes these two types of dynamics at
  time $t$ based on flipping a biased coin: The position $x_{t+1}$ of
  the particle at the next time $t+1$ is determined by choosing with
  probability $1-p$ the diffusive system $D$ while $L$ is picked with
  probability $p$.  The trajectory generated by $x_{t+1}=R(x_t)$
  displays {\em intermittency}, where long regular phases alternate
  randomly with irregular-looking, chaotic motion.}
\label{fig:model}
\end{figure}

\begin{figure}[htbp]
  \centering
\includegraphics[scale=0.65]{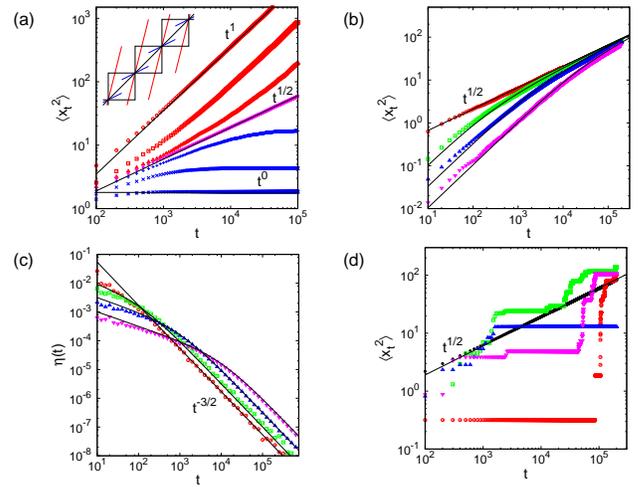}

\caption{Mean square displacement (MSD) and waiting time distribution
  (WTD) for randomized deterministic diffusion.  The two deterministic
  dynamical systems that are randomly sampled in time with probability
  $p$ by applying the recipe of Fig.~\ref{fig:model} are illustrated
  in the inset of (a).  All symbols are generated from computer
  simulations. (a) MSD $\langle x_t^2\rangle$ for
  $p=0.6,0.63,0.663,2/3,0.669,0.68,0.7$ (top to bottom) for an
  ensemble of particles, where each particle experiences a different
  random sequence.  There is a characteristic transition between
  normal diffusion and localization via subdiffusion at a critical
  $p_c=2/3$. (b) MSD at $p_c$ by starting the computations after
  different ageing times $t_a=0,10^2,10^3,10^4$ (top to bottom). The
  MSD displays ageing similar to analytical results from continuous
  time random walk (CTRW) theory \cite{Bar03} (bold lines). (c) WTD
  $\eta(t)$ at $p_c$ for particles leaving a unit interval at the same
  ageing times $t_a$ as in (b). The bold lines are again analytical
  results from CTRW theory \cite{Bar03}.  (d) MSD at the critical
  probability $p_c$ for different types of averaging over the random
  variable. For the straight black line with matching symbols each
  particle experiences a {\em different} random sequence (called
  uncommon noise), cf.\ Fig.~\ref{fig:msdandageing}(a).
  The other four lines depict MSDs obtained from applying the {\em
    same} sequence of random variables to all particles (called common
  noise). In these four cases the MSD becomes a random variable
  breaking self-averaging.}

\label{fig:msdandageing}
\end{figure}

Figure~\ref{fig:model} gives our recipe to combine two deterministic
dynamical systems $D$ and $L$ randomly in time. Here $D$ generates
normal diffusion while $L$ yields localization of particles.  We
sample randomly between both systems with probability $p$ of choosing
$L$ at discrete time step $t\in\mathbb{N}$, respectively probability
$1-p$ of chooosing $D$.  For $p=0$ we thus recover the dynamics of $D$
while for $p=1$ we obtain the one of $L$. This implies that there must
exist a transition between these two different dynamics under
variation of $p$. Our central question is: For $0<p<1$, what type of
diffusive dynamics emerges in the resulting {\em random dynamical
  system} $R$? Here we model deterministic diffusion by chaotic random
walks on the line \cite{GF2,GeNi82,SFK,Kla06} defined by the equation
of motion $x_{t+1}=M_a(x_t)$, where
\begin{equation}
M_a(x)  = \left\{ 
\begin{array}{r@{\quad,\quad}l}
a x & 0\le x<\frac{1}{2} \\ 
a x +1-a & \frac{1}{2} \le x< 1
\end{array}
\right. \: , \: a>0 \label{eq:mapa} \: ,
\end{equation}
is a piecewise linear map lifted onto $\mathbb{R}$ by
$M_a(x+1)=M_a(x)+1$, cf.\ the inset in Fig.~\ref{fig:msdandageing}(a).
For $a>2$ this model exhibits normal diffusion with a Lyapunov
exponent calculated to (see Sec.~1 in our Supplement \cite{suppl},
which includes Refs.~\cite{Ott,Rob95,ASY97,Do99,dslnotes,Kla17})
$\lambda(a)=\ln a$ \cite{RKD,KlDo99,GrKl02,Kla06}. The sample
trajectory in the upper left of Fig.~\ref{fig:model} was obtained from
$D=M_4(x)$, where the dynamics is chaotic according to $\lambda(4)=\ln
4>0$. The trajectory in the upper right of Fig.~\ref{fig:model}
corresponds to $L=M_{1/2}(x)$, where the dynamics is non-chaotic due
to $\lambda(1/2)=-\ln 2<0$. Here all particles contract onto stable
fixed points at integer positions $x\in\mathbb{Z}$. For defining the
{\em random map} $R$ the slope $a$ becomes an independent and
identically distributed, multiplicative random variable: At any time
step $t$ we choose for our map $R=M_a(x)$ with probability $p\in[0,1]$
the slope $a=1/2$ while with probability $1-p$ we pick $a=4$. The
sequence of random slopes may or may not depend on the individual
particle if we consider an ensemble of them \cite{BAE13}, as we
explore below. Random maps of this type are also called iterated
function systems \cite{Barn88,AGH18}. They have been studied by
both mathematicians and physicists in view of their measure-theoretic
\cite{Pel84,AAN98,AGH18} and statistical physical properties
\cite{LOC90,RKla01b,LBDF04,BAE13}.

\begin{figure*}[htbp]
\centering
    \includegraphics[scale=0.85]{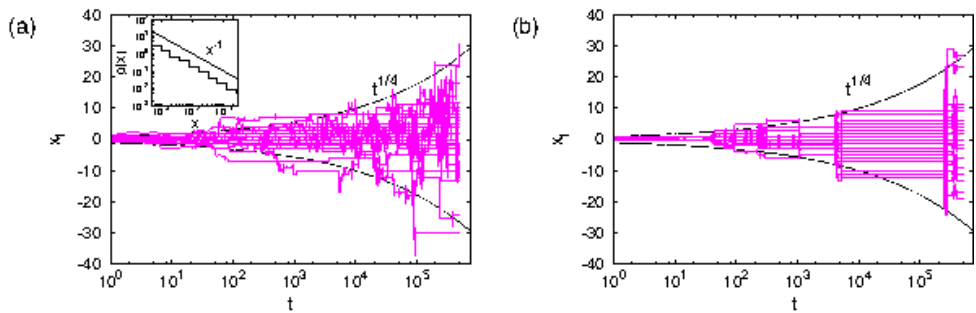}
    \caption{Subdiffusion for different types of randomness.  (a)
      Sample trajectories at $p_c$ corresponding to 30 different
      initial conditions with uncommon noise. (b) Same as (a) with
      common noise. The envelopes in (a) and (b) correspond to the
      subdiffusive spreading for uncommon noise shown in
      Fig.~\ref{fig:msdandageing}(d).  (b) displays jump time
      synchronization of all particles. The inset in (a) yields in
      double-logarithmic plot the infinite invariant density within a
      unit cell for uncommon noise for the map $R \; \mbox{mod 1}$.}
\label{fig:ergbreak}
\end{figure*}

One can show straightforwardly that the Lyapunov exponent
  $\lambda(p)$ of the random map $R$ is zero at probability $p_c=2/3$
  \cite{suppl}. Since $\lambda(p)>0$ for $p<p_c$ the map $R$ should
generate normal diffusion in this regime while $p>p_c$ with
$\lambda(p)<0$ should lead to localization for long times.  In
Fig.~\ref{fig:msdandageing} we test this conjecture by comparing
numerical with analytical results.  For our simulations we used $\sim
10^5$ iterations of $R$ with $\sim 10^5$ initial points, which were
distributed randomly and uniformly in the unit interval $[0,1)$. Here
  each particle experienced a different sequence of random slopes. We
  used an arbitrary precision algorithm with up to $10^{10000}$
  decimal digits.  Figure~\ref{fig:msdandageing}(a) depicts the MSD
  $\langle x^2_t\rangle$ under variation of $p$ by confirming the
  diffusion scenario conjectured above. However, passing through $p_c$
  the dynamics displays a subtle transition: Right at $p_c$ we obtain
  long-time subdiffusion, $\langle x^2(t)\rangle\sim t^{1/2}$, while
  around $p_c$ this dynamics survives for long transient times.
  Figures~\ref{fig:msdandageing}(b) and (c) reveal that right at $p_c$
  $R$ exhibits {\em ageing} \cite{Bou92,Metz15} in both the MSD
  and the waiting time distribution (WTD). The latter is the
  probability distribution $\eta(t)$ of the times $t$ it takes a
  particle to escape from a unit interval of $R$. In both (b) and (c)
  there is good agreement with analytical results from continuous time
  random walk (CTRW) theory for long times, $\langle x_t^2\rangle\sim
  (t+t_a)^{\alpha}-t_a^{\alpha}$ and $\eta(t)\sim
  t_a^{\alpha}/[(t+t_a)t^{\alpha}]$, where $t_a$ is the ageing time
  \cite{Bar03}. This theory furthermore predicts that for long times a
  WTD of $\eta(t)\sim t^{-\gamma}$ implies a MSD of $\langle
  x^2(t)\rangle\sim t^{\gamma-1}$
  \cite{GeTo84,ZuKl93a,Bar03,KCKSG06,KKCSG06}. For $R$ this relation
  is fulfilled with $\gamma=3/2$. An exponent of the WTD of $3/2$
    yields a diverging mean waiting time. This as well as the
    existence of ageing imply {\em weak ergodicity breaking} of the
    dynamics \cite{Bou92,BeBa05,Metz15}.

However, our map $R$ generates dynamics that goes beyond conventional
CTRW theory. This becomes apparent by looking at different types of
averaging over the random variables shown in
Fig.~\ref{fig:msdandageing}(d): While in
Fig.~\ref{fig:msdandageing}(a)-(c) each particle experienced a {\em
  different} sequence of random slopes, as reproduced by the straight
line with matching symbols for the MSD in
Fig.~\ref{fig:msdandageing}(d), for all the other MSDs in (d) the
corresponding random sequences are the {\em same} for all particles.
Accordingly, we call the former type of randomness {\em uncommon
  noise}, the latter {\em common noise}.  Crucially, while in
Fig.~\ref{fig:msdandageing}(a), based on uncommon noise, the MSD is
well-defined for all $p$, Fig.~\ref{fig:msdandageing}(d) shows that
for common noise sequences it becomes a random variable at $p_c$
in the long time limit that completely depends on the random
squence chosen. This bears strong similarity to what is called {\em
  breaking of self-averaging} for random walks in quenched disordered
environments \cite{BoGe90}, which also implies weak ergodicity
breaking \cite{DRG16}.

Figure~\ref{fig:ergbreak} displays space-time plots of 30 trajectories
starting at different initial points for (a) uncommon noise and (b)
common noise. While in (a) the different trajectories look rather
irregular yielding a spreading front that matches to the subdiffusion
depicted in Fig.~\ref{fig:msdandageing}(d) for uncommon noise, (b)
shows `temporal clustering' in the form of jump time synchronization,
i.e., all particles eventually jump from unit cell to unit cell at the
same time step. This matches to the fact that the MSD does not
converge for common noise as seen in Fig.~\ref{fig:msdandageing}(d).
The inset in (a) represents the invariant density of the map $R \;
\mbox{mod 1}$, i.e., within a unit cell, with uncommon noise. We see
that it decays on average like $\rho(x)\sim x^{-1}$. This result and
the stepwise structure of $\rho(x)$ are in agreement with analytical
calculations \cite{Pel84,AAN98}. At zero Lyapunov exponent uncommon
noise thus leads to a weak spatial clustering \cite{WMGW12} and path
coalescence \cite{WiMe03} of particles at integer positions
$x\in\mathbb{Z}$. In contrast, for common noise an invariant density
does not exist, and we do not find any spatial clustering.

We now explore the origin of this type of anomalous dynamics in terms
of dynamical systems theory. As exemplified by the trajectory shown at
the bottom of Fig.~\ref{fig:model}, around $p_c$ the dynamics of $R$
is {\em intermittent} \cite{BLMV11} meaning long regular phases
alternate randomly with irregular-looking, chaotic motion. A
paradigmatic intermittent dynamical system is the Pomeau-Manneville
map $P_{z,b}(x)=x+bx^z\:,\:b\ge1\:,\:x\in[-1/2,1/2)$. Defining its
  equation of motion in the same way as for $M_a$ above it generates
  subdiffusion characterized by a MSD and a WTD that in suitable
  scaling limits match to the predictions of CTRW theory with
  $\gamma=z/(z-1)$ \cite{GeTo84,ZuKl93a,KKCSG06}. As shown above, for
  the map $R$ CTRW theory correctly predicts the relation between the
  long-time MSD and the WTD by using $\gamma=3/2$. Trying to
  understand the random map $R$ in terms of $P_{z,b}$ thus suggests to
  choose $z=3$. One should now compare the invariant densities
  $\rho(x)$ of the two maps $\mbox{mod 1}$: For $P_{z,b} \; \mbox{mod
    1}$ it is known that $\rho(x)\sim x^{1-z}\:,\:x\gg0$, which for
  $z\ge2$ becomes a non-normalizable, infinite invariant density
  \cite{Tha83,KoBa09}. But for $z=3$ this yields $\rho(x)\sim
  x^{-2}$ while for $R\;\mbox{mod 1}$ we have $\rho(x)\sim x^{-1}$,
  see the inset of Fig.~\ref{fig:ergbreak}(b) \cite{iidens}. Hence,
  the intermittency displayed by $R$ is not of Pomeau-Manneville type
  but of a fundamentally different microscopic dynamical origin. This
  might relate to deviations between CTRW theory, which on a
  coarse-grained level works well for the Pomeau-Manneville map, and
  our numerical results for $R$ on short time scales in the MSD and
  the WTD of Fig.~\ref{fig:msdandageing}. It would be interesting
    to further explore such differences, e.g., by the approach
    outlined in Ref.~\cite{AlRa13}.

However, there is another type of intermittency in dynamical systems
that is profoundly different from Pomeau-Manneville dynamics, called
{\em on-off intermittency}
\cite{Pik84,FuYa85,FuYa86,PiGr91,PST93,HPH94}. It was first reported
for two-dimensional coupled maps
\begin{eqnarray}
x_{t+1}&=&(1-\epsilon)f(x_t)+\epsilon f(y_t)\nonumber\\
y_{t+1}&=&(1-\epsilon)f(y_t)+\epsilon f(x_t)\quad ,
\label{eq:coma}
\end{eqnarray}
where $x_{t+1}=f(x_t)$ is chaotic with positive Lyapunov exponent and
$\epsilon\in[0,1]$ \cite{Pik84,FuYa85}. When $\epsilon$ is large, the
possibly chaotic dynamics is trapped on the synchronization manifold
$x_t=y_t$. By decreasing $\epsilon$ to a critical parameter
$\epsilon=\epsilon^*$ trajectories start to escape from this manifold
into the full two-dimensional space. This is called a {\em
  blowout bifurcation} and the associated intermittency on-off
intermittency \cite{OtSo94}. In subsequent works Eqs.~(\ref{eq:coma})
were boiled down to more specific two-dimensional maps
\cite{Pik84,PiGr91,FuYa86,LOC90,PST93,HPH94,HaMi97,AAN98}. The
simplest ones are piecewise linear \cite{PST93,HaMi97,AAN98}, such as
\cite{AAN98}
\begin{eqnarray}
  x_{t+1}&=&\left\{
  \begin{array}{cc}
    ax_t& (x_t<1\:,\:0\le y_t \le p)\\
    \frac1a x_t& (x_t<1\:,\:p<y_t\le 1)\\
    1+b(1-x_t)& (x_t\ge1) 
  \end{array}
  \right.\nonumber\\
  y_{t+1}&=&\left\{
    \begin{array}{cc}
    \frac{y_t}{p}& (0\le y_t\le p)\\
    \frac{y_t-p}{1-p}& (p<y_t\le 1) \quad ,
    \end{array}
    \right. 
    \label{eq:rlm}
\end{eqnarray}
with symmetry $y\to-y$ and parameters $a>0,b\in\mathbb{R},p\in(0,1)$.
Due to its skew product form this system can be understood as a
one-dimensional map $x_{t+1}=f(x_t)$ with multiplicative randomness
generated by $y_{t+1}=g(y_t)$
\cite{Pik84,PiGr91,FuYa86,HaMi97,LOC90,HPH94}. In a next step one
might replace the deterministic chaotic dynamics of $y_t$ by
stochastic noise. If we now consider the dynamics of $x_t$ in
Eqs.~(\ref{eq:rlm}) on the unit interval only by choosing $a=2$,
taking the map $\mbox{mod 1}$ and choosing dichotomic noise, we obtain
a simple piecewise linear map with multiplicative randomness that is
qualitatively identical to our model $R \;\mbox{mod 1}$
\cite{Pel84,LOC90,AAN98}.  For this class of systems it has been shown
numerically and analytically that at a critical $p_c$ the invariant
density of $x=x_t$ decays like $\rho(x)\sim x^{-1}$
\cite{PiGr91,FuYa85,HPH94,HaMi97,AAN98,HHF99,MHB01} and that a
suitably defined waiting time distribution between chaotic `bursts'
obeys $\eta(t)\sim t^{-3/2}$ \cite{HPH94,HaMi97,AAN98,HHF99,MHB01}. In
Refs.~\cite{HHF99,MHB01} different diffusive models driven by on-off
intermittency have been studied, and for two of them \cite{MHB01}
subdiffusion with a MSD of $\langle x^2(t)\rangle\sim t^{1/2}$ has
been obtained by matching simulation results to CTRW theory. We thus
conclude that our model $R$ exhibits anomalous diffusion generated by
on-off intermittency. We emphasize, however, that the mechanism
underlying our model depicted in Fig.~\ref{fig:model} is more general
than this particular type of intermittent dynamics.

In order to check for the generality of our results, in the Supplement
\cite{suppl} we first replace the dichotomic noise by physically more
realistic continuous noise distributions choosing 1.\ uniform noise on
a bounded interval, and 2.\ a non-uniform unbounded log-normal
distribution. Figures~1 and 2 in Secs.~2 and 3 of \cite{suppl},
respectively, show that our mechanism is very robust under variation
of the type of noise. We may thus conjecture that our scenario of
subdiffusion generated by random maps holds for any generic type of
noise. We also tested whether the strong localisation due to
contraction onto a stable fixed point can be replaced by a weaker
chaotic localisation to a subregion in phase space. However, in this
case the transition between diffusive and localised dynamics is
entirely different without displaying any subdiffusion, cf.\ Fig.~3 in
Sec.~4 of \cite{suppl}. As a general principle, one must thus mix
expansion with contraction to generate anomalous dynamics. Finally, in
Sec.~5 of \cite{suppl} we study a simple nonlinear map that exhibits
different types of diffusion in different parameter regions of a
bifurcation scenario generating chaotic and periodic windows.
Randomizing this map according to Fig.~\ref{fig:model} yields again
subdiffusion with a MSD of $\langle x_t^2\rangle\sim t^{1/2}$ and a
WTD of $\eta(t)\sim t^{-3/2}$, cf.\ Fig.~4 in \cite{suppl} which
includes Refs~\cite{BBJ85,KoKl02,KoKl03}. This demonstrates that the
basic mechanism generating anomalous diffusion which we propose is
also robust in a nonlinear setting.

In summary, we have shown that anomalous dynamics emerges if we
randomly mix chaotic diffusion and non-chaotic localisation with a
sampling probability yielding a zero Lyapunov exponent of the
randomized dynamics. Interestingly, our basic mechanism bears
similarity with the famous problem of a protein searching for a target
at a DNA strand \cite{BWvH81}: Here the protein randomly switches
between (normal) diffusion in the bulk of the cell and moving along
the DNA. This is called {\em facilitated diffusion}, as the random
switching between different modes may decrease the average time to
find a target \cite{BWvH81,MEK09,BaMe12}. We are not aware, however,
that for this problem any emergence of anomalous diffusion as an
effective dynamics representing the whole diffusion process has been
discussed. Along these lines, one might speculate that using our
framework for combining normal diffusion with constant velocity
scanning \cite{MEK09} could yield a kind of L\'evy walk \cite{ZDK15},
which poses an interesting open problem.

\begin{acknowledgments}
Y.S.\ is funded by the Grant in Aid for Scientific Research (C) No.
18K03441, JSPS, Japan. R.K.\ thanks Prof.~Krug from the U.\ of Cologne
and Profs.~Klapp and Stark from the TU Berlin for hospitality as a
guest scientist as well as the Office of Naval Research Global for
financial support.  Y.S.\ and R.K.\ acknowledge funding from the
London Mathematical Laboratory, where they are External Fellows, and
thank two anonymous referees for very helpful comments.
\end{acknowledgments}

\newpage

\setcounter{equation}{0}
\setcounter{figure}{0}
\setcounter{table}{0}
\setcounter{page}{1}
\renewcommand{\figurename}{Supplementary Figure}
\renewcommand{\tablename}{Supplementary Table}
\renewcommand{\theequation}{S\arabic{equation}}
\renewcommand{\thetable}{S\arabic{table}}
\renewcommand{\thefigure}{S\arabic{figure}}

\begin{appendix}

\onecolumngrid

\vspace*{0.5cm}
\centerline{\Large \bf Supplemental Material}

\section{1. Calculation of the Lyapunov exponent for random maps}

For one-dimensional maps $M(x)$ obeying the deterministic equation of
motion $x_{t+1}=M(x_t)$ the {\em local Lyapunov exponent} is defined by
\cite{Ott,Rob95,ASY97,Do99,dslnotes,Kla17}
\be
\lambda(x)=\lim_{t\to\infty} \frac{1}{t}\sum_{k=0}^{t-1} 
\ln|M'(x_k)|\quad ,
\label{eq_lyap}
\ee
where $M'(x)$ denotes the derivative of the map. By definition here
the Lyapunov exponent $\lambda$ depends on the initial condition
$x=x_0$ of the map, hence it is called local. Eq.~(\ref{eq_lyap})
represents a time average along the trajectory
$\{x_0,x_1,\ldots,x_k,\ldots,x_{t-1}\}$ of the map. If the map is
ergodic the dependence on initial conditions will disappear
\cite{Do99,dslnotes,Kla17}. For the piecewise linear map $M_a(x)$ defined
by Eq.~(1) in the main text we have that $M_a'(x)=a$, hence
$\lambda(a)=\ln a$ as noted in the main text on p.1.

We now apply Eq.~(\ref{eq_lyap}) to calculate the Lyapunov exponent for
our random maps $R=M_a(x)$ as defined on p.1 and 2 in the main text,
where $M_a(x)$ is again the piecewise linear map given by Eq.~(1) in the
main text. Therein the slope $a$ becomes a random variable drawn from
a probability distribution $\chi(\xi)\:,\:a=\xi$, at any time step $t$,
$a=a_t$. We first choose for $\chi$ dichotomic noise,
where $a$ is sampled independently and identically distributed with
probability $p\in[0,1]$ from $a=a_{loc}$ and with probability $1-p$
from $a=a_{exp}$. Feeding this information into Eq.~(\ref{eq_lyap}) we
obtain \bna \lambda_{dich}&=&\lim_{t\to\infty}
\left(\frac{t_{loc}}{t}\ln a_{loc}+\frac{t_{exp}}{t}\ln
a_{exp}\right)\nonumber\\ &=&p\ln a_{loc}+(1-p)\ln a_{exp} \quad ,
\label{eq_ldich}
\ena
where $t_{loc}$ and $t_{exp}$ yield respective numbers of events out of
a total of $t$ time steps for which the particle experiences the
contracting and thus localising, respectively the expanding map
\cite{Do99}. In the long time limit these fractions are identical with
the corresponding sampling probabilities $p$ and $1-p$ leading to our
final result. Note that in our piecewise linear map $R=M_a(x)$ all
particles are exposed to the same uniform slope $a$ irrespective of
initial conditions, hence $\lambda_{dich}=\lambda_{dich}(x)$.

From this equation we now calculate the critical sampling probability
$p_c$ at which $\lambda_{dich}=0$ by choosing $a_{loc}=1/2$ and
$a_{exp}=4$ as in the main text, which defines our first model. It follows 
\be
0=p\ln 1/2+(1-p)\ln 4\quad ,
\ee
which is solved to $p_c=2/3$ as stated on p.2 in the main text.

By observing that dichotomic noise is defined by the probability distribution
\be
\chi_{dich}(\xi)=p\delta(\xi-a_{loc})+(1-p)\delta(\xi-a_{exp})
\label{eq_dich}
\ee
we can rewrite Eq.~(\ref{eq_ldich}) more generally as
\be
\lambda_{dich}=\int_{a_{loc}-\epsilon}^{a_{exp}+\epsilon}d\xi\:\chi_{dich}(\xi)\ln|\xi|
\label{eq_ldichpdf}
\ee
with $0<\epsilon\ll1$. From the above line of arguments it
follows that for an arbitrary noise distribution $\chi(\xi)\:,\:\xi\in I$
we can calculate the Lyapunov exponent $\lambda_{\chi}$ of a random dynamical system $x_{t+1}=R(\xi_t,x_t)$ with map
\bna
  R(\xi,x)=\left\{
  \begin{array}{ll}
    \xi x & 0\le x<\frac12\\
    \xi(x-1)+1 & \frac12\le x<1\\
  \end{array}
  \right.
  \label{eq_mapr}
\ena
lifted onto the whole real line according to $R(x+1)=R(x)+1$ by
\be
\lambda_{\chi}=\int_Id\xi\:\chi(\xi)\ln|\xi|
\label{eq_lchi} \quad .
\ee
Equations~(\ref{eq_ldichpdf}),(\ref{eq_lchi}) thus express
the Lyapunov exponent of a random map $R$ by ensemble averages over
the corresponding noise distributions instead of time averages along
trajectories, which more generally presupposes ergodicity of the dynamics
\cite{Do99,dslnotes,Kla17}. For maps that are more complicated than
our piecewise linear case $M_a(x)$ Eq.~(1) in the main text,
calculating $\lambda_{\chi}$ will require integration over the full
invariant density $\rho(x,\eta)$ of the random map, which can be a
very non-trivial object \cite{SDLR19}.

\section{2. Random map with uniform distribution of slopes}

We now consider a second model that is more general than the one
defined by dichotomic noise. For this purpose we sample the slopes of
the piecewise linear map Eq.~(\ref{eq_mapr}) from noise $\chi(\xi)$
uniformly distributed over an interval $\xi\in I=[a,b]\:,a>0$, that
is, we choose

\be
\xi\sim \chi_{unif}(\xi)=\frac{1}{b-a} \quad .
\label{eq_unif}
\ee
For this model the Lyapunov exponent can be calculated from
Eq.~(\ref{eq_lchi}) to
\be
\lambda_{unif}=\int_a^{b} d\xi\:\frac{1}{b-a} \ln\xi =b\log b -b -(a\log a-a)
\quad .
\ee

\begin{figure}[htbp]
  \includegraphics[width=0.4\textwidth]{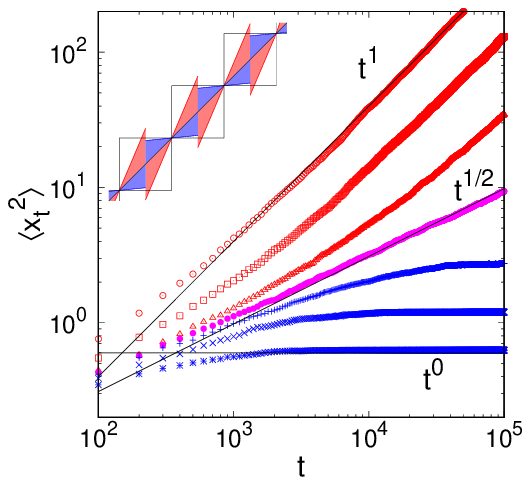}
  \includegraphics[width=0.55\textwidth]{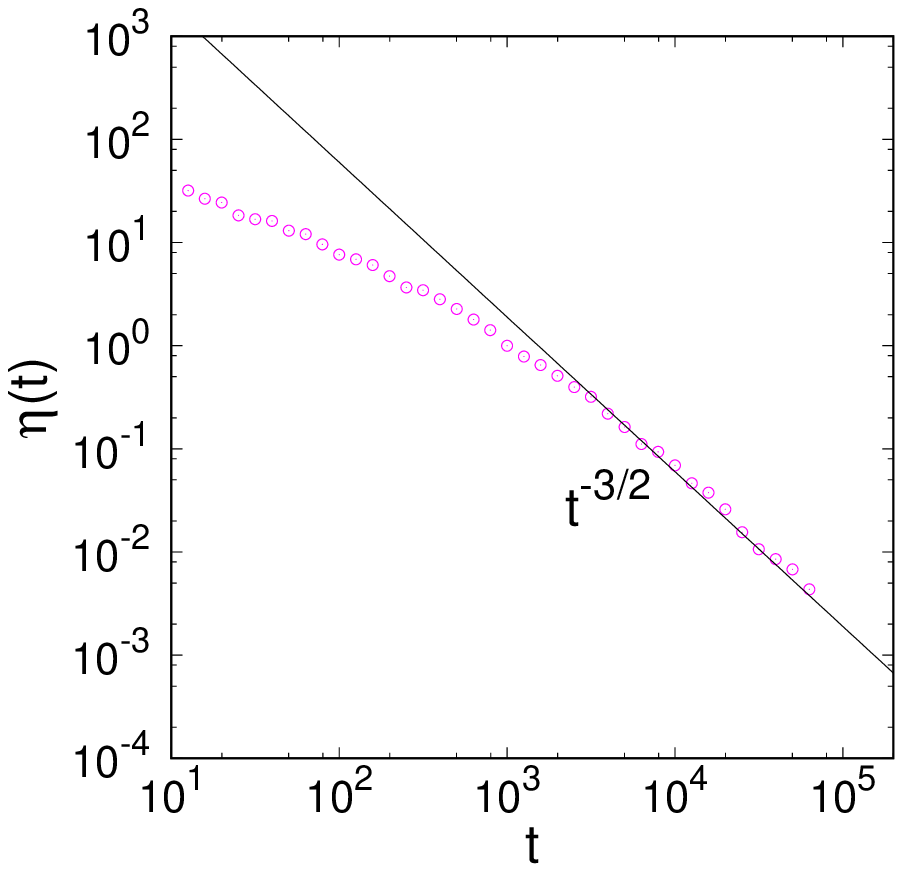}
\caption{Mean square displacement (MSD) and waiting time distribution
  (WTD) for the random map Eq.~(\ref{eq_mapr}) where the slopes are
  sampled from the uniform noise distribution Eq.~(\ref{eq_unif}); see
  the inset on the left for an illustration. All symbols in the plots
  are generated from computer simulations. (left) MSD $\langle
  x_t^2\rangle$ for $b =2.5, 2.45, 2.38, 2.36384, 2.35, 2.33, 2.3$
  (top to bottom) for an ensemble of particles, where each particle
  experiences a different random sequence. In complete analogy to
  Fig.~2(a) of the main text there is a characteristic transition
  between normal diffusion and localization via subdiffusion at a
  critical $b_c\simeq 2.36384$. (right) The WTD $\eta(t)$ at
  $b_c\simeq 2.36384$ showing a power law with exponent $-3/2$,
  cp.\ with Fig.~2(c) in the main text.}
\label{msd_uniform_noise}
\end{figure}

Choosing $a=0.1$ and keeping it fixed by varying the upper bound $b$,
we obtain the critical parameter $b_c$ at which $\lambda_{unif}=0$ to
$b_c\simeq 2.36384$. As for dichotomic noise we expect the resulting
dynamics at $b_c$ to be subdiffusive, which is confirmed in
Fig.~\ref{msd_uniform_noise}. We conclude that our main result of
subdiffusion in a random map is robust against generalising the noise
distribution from a dichotomic one to uniform noise on a bounded
interval.

\section{3. Random map with log-normal distribution of slopes}

\begin{figure}[htbp]
  \includegraphics[width=0.4\textwidth]{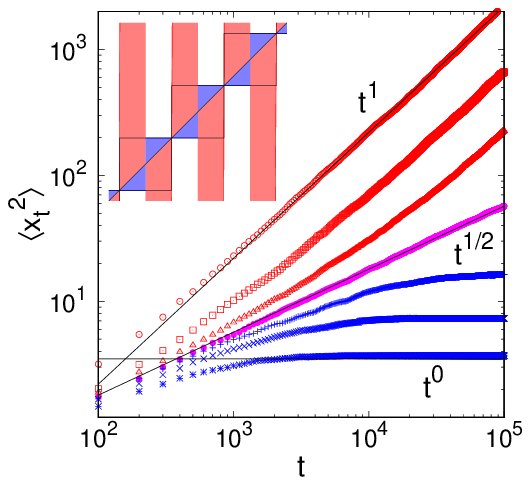}
  \includegraphics[width=0.55\textwidth]{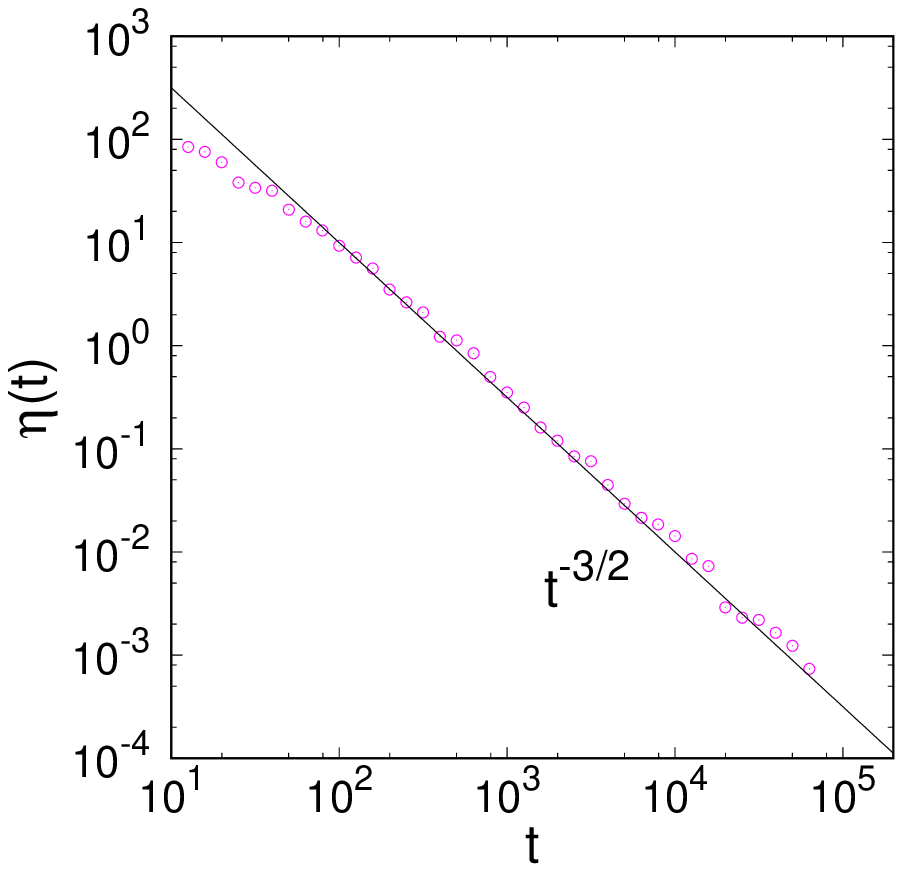}
  \caption{MSD and WTD for the random map Eq.~(\ref{eq_mapr}) where
    the slopes are sampled from the log-normal distribution
    Eq.~(\ref{eq_logn}); see the inset on the left for an
    illustration. All symbols in the plots are generated from computer
    simulations. (left) MSD $\langle x_t^2\rangle$ for $\mu=0.1, 0.03,
    0.01, 0, -0.07, -0.015, -0.03=$ (top to bottom) for an ensemble of
    particles, where each particle experiences a different random
    sequence. In complete analogy to Fig.~2(a) of the main text there
    is a characteristic transition between normal diffusion and
    localization via subdiffusion at a critical $\mu=0$.  (right) The
    WTD $\eta(t)$ at $\mu=0$ showing a power law with exponent $-3/2$,
    cp.\ with Fig.~2(c) in the main text.}
\label{msd_logn}
\end{figure}

Our third model is defined by sampling the slopes in the random map
Eq.~(\ref{eq_mapr}) from a log-normal distribution,
\be
\xi\sim \chi_{logn}(\xi)=\mbox{Lognormal($\mu$,1)}=\frac{1}{\xi\sqrt{2\pi}}\exp\left(-\frac{(\ln \xi -\mu)^2}{2}\right) \quad .
  \label{eq_logn}
\ee
The Lyapunov exponent can again be calculated from
Eq.~(\ref{eq_lchi}) to
\be
\lambda_{logn}=\int_0^{\infty}d\xi\: \frac{1}{\xi\sqrt{2\pi}}\exp\left(-\frac{(\ln \xi -\mu)^2}{2}\right) \ln\xi=\mu.
\ee
For $\lambda_{logn}=0$ this yields trivially a critical parameter
value of $\mu_c=0$ at which we expect the random map to generate
subdiffusion. This is indeed confirmed in Fig.~\ref{msd_logn}.  We
conclude that our main result of subdiffusion in a random map is also
robust against generalising the noise distribution from uniform noise
on a bounded interval to non-uniform noise on an unbounded interval.

\section{4. Random map with chaotic trapping}

We now investigate whether the vanishing of the Lyapunov exponent
exploited in the examples studied above is really necessary for having
subdiffusion in our random maps. For this purpose we introduce a
fourth model which is similar to our very first model
Eqs.~(\ref{eq_dich}),(\ref{eq_mapr}) but for which we choose two
slopes at which the corresponding deterministic maps are both
expanding, $a_{loc}=3/2$ and $a_{exp}=4$. Note that for $a_{loc}=3/2$
the deterministic map Eq.~(1) in the main text has a positive Lyapunov
exponent $\lambda(3/2)= \ln 3/2$, see our calculation at the beginning
of Sec.~1. Nevertheless, this map does not generate any long-time
diffusion, since particles cannot escape from any unit interval due to
the fact that the map does not exceed the unit interval. Hence these
sets are trivially decoupled by generating chaotic trapping of
particles on any unit interval.

\begin{figure}[htbp]
  \includegraphics[width=0.48\textwidth]{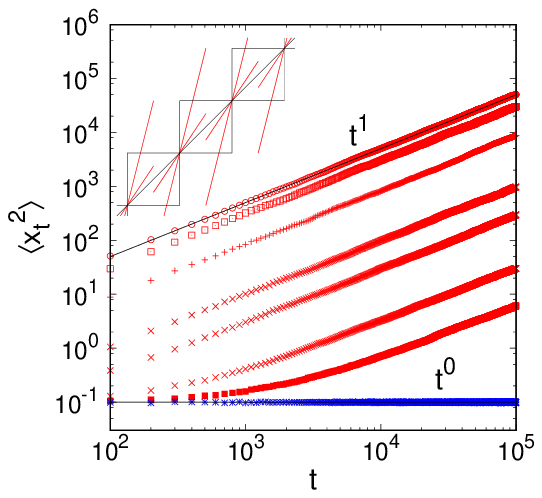}
  \includegraphics[width=0.35\textwidth]{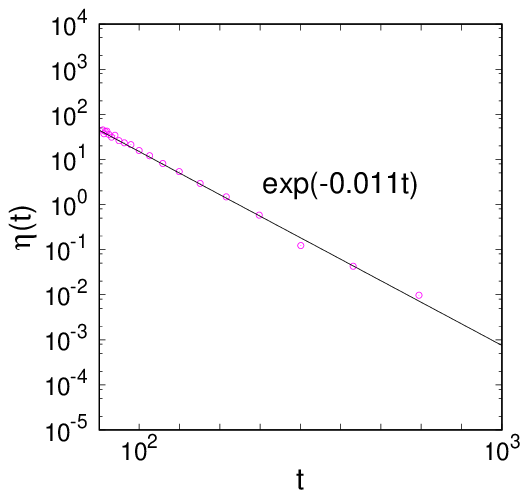}
  \caption{MSD and WTD for the random map Eq.~(\ref{eq_mapr}) where
    the slopes are sampled from the dichotomic distribution
    Eq.~(\ref{eq_dich}) with slopes $a_{loc}=3/2$ and $a_{exp}=4$; see
    the inset on the left for an illustration. Thus, in contrast to
    the map used for the corresponding results presented in Fig.~2 of
    the main text here the localisation is not due to contraction onto
    a stable fixed point but the dynamics remains chaotic on a bounded
    interval; see the text for further explanations. (left) MSD
    $\langle x_t^2\rangle$ for $p =0.0, 0.5, 0.9, 0.99, 0.995, 0.9995,
    0.9999, 1.0$ (top to bottom) for an ensemble of particles, where
    each particle experiences a different random sequence. In marked
    contrast to Fig.~2 of the main text, Fig.~\ref{msd_uniform_noise}
    and Fig.~\ref{msd_logn} above there is no characteristic
    transition between normal diffusion and localization via
    subdiffusion. (right) The WTD $\eta(t)$ at $p=0.99$ showing an
    exponential distribution, which is again in sharp contrast to the
    corresponding power law WTDs in the three figures mentioned
    above.}
\label{msd_chaotic_trapping}
\end{figure}

Figure~\ref{msd_chaotic_trapping} demonstrates that for this purely
chaotic model there is no transition between normal diffusion and
localisation via subdiffusion. In marked contrast to all our previous
models we see that for $p>0$ the map always generates normal diffusion
in the long time limit. Instead of subdiffusion the MSD displays
longer and longer transients for shorter times at which diffusion is
more and more slowed down when approaching the strictly localised
dynamics at $p=1$. This numerical finding is confirmed by a WTD that
is exponential even very close to the localisation value at $p=1$ as
also shown in Fig.~\ref{msd_chaotic_trapping}. It is in fact well
known that for chaotic dynamical systems characterised by a positive
Lyapunov exponent the WTD is always exponential while for non-chaotic
systems exhibiting regular dynamics it decays as a power law
\cite{BaBe90,Gasp}. For an exact analytical calculation of the
exponential WTD in an open version of the deterministic map Eq.~(1)
(main text) we refer to Refs.~\cite{Ott,Do99,Kla17}. To explicitly
calculate the WTDs for this and our other random maps, possibly along
these lines, in order to quantitatively confirm our numerical results
from first principles remains an interesting open problem.  We thus
conclude that the condition of having a zero Lyapunov exponent is
indispensable for observing subdiffusion in these random maps. This is
deeply rooted in a mechanism yielding anomalous diffusion that
requires an intricate interplay between contracting and expanding
dynamics mixed in time.

\section{5. Random nonlinear map}

\begin{figure}[htbp]
  \centering
  \includegraphics[scale=1.0]{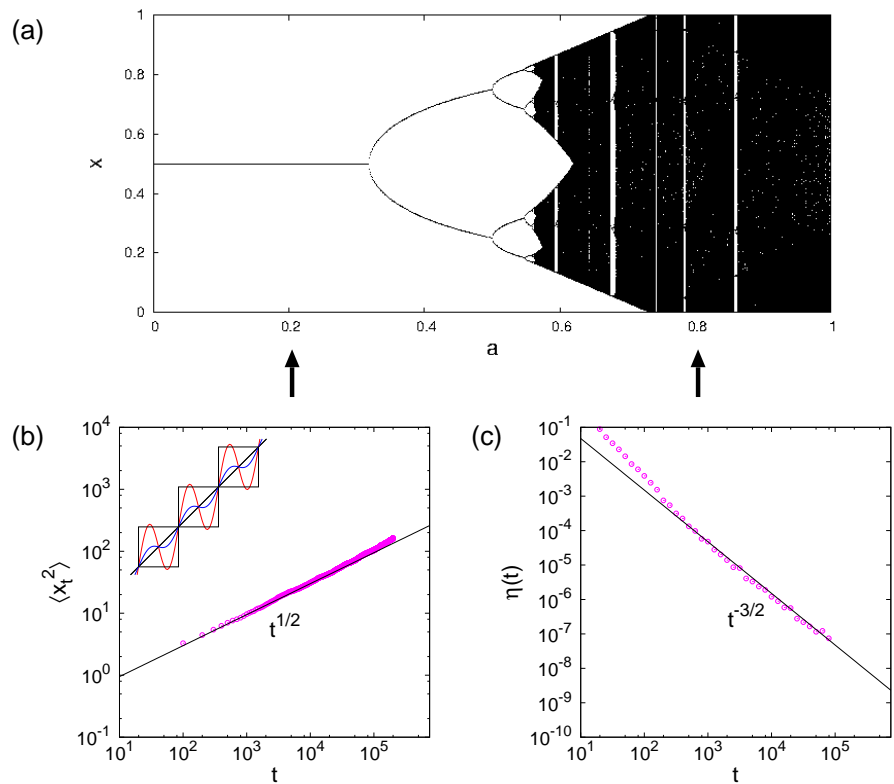}
  \caption{Subdiffusion in the random climbing sine map.  (a) The
    bifurcation diagram of the climbing sine map depicted in the inset
    of (b) \cite{KoKl02,KoKl03}.  The two arrows denote the two
    specific parameter values $c_1=0.8$ and $c_2=0.2$ chosen with
    probability $1-p$, respectively $p$, to generate a random
    dynamical system from this map as explained in the text. (b) MSD
    at the critical probability $p_c\simeq 0.505702$ where the random
    map's Lyapunov exponent is approximately zero.  The symbols are
    from simulations, the line is a fit based on CTRW theory. The MSD
    clearly exhibits subdiffusion with exponent $1/2$.  (c) The WTD
    for the same $p_c$ showing a power law with exponent $-3/2$ as
    predicted by CTRW theory; symbols and lines are as in (b).}
\label{fig:csine}
\end{figure}

We finally test whether the observed subdiffusion in random
maps also persists in nonlinear maps. As a fifth model we thus replace
the map $M_a(x)$ Eq.~(1) in the main text by the climbing sine map,
\be
C_c(x)=x+c\sin(2\pi x)\:,\:x\in\mathbb{R}\quad,
\ee
sketched in the inset of Fig.~\ref{fig:csine}(b) \cite{SFK}. This map
can be derived from discretizing a driven nonlinear pendulum equation
in time \cite{BBJ85} thus representing a much more general class of
nonlinear dynamics than piecewise linear maps. It was found that $C_c$
displays three different regimes of diffusion under parameter
variation with an exponent of the MSD of $\alpha=0$, $1$ or $2$
\cite{KoKl02,KoKl03}.  These regimes correspond to different parameter
regions in the map's bifurcation diagram reproduced in
Fig.~\ref{fig:csine}(a): $\alpha=1$ occurs in the chaotic regions
while $\alpha=0$ and $2$ match to two different classes of periodic
windows, where all particles either converge onto attracting localized
periodic orbits, or onto ballistic ones.

Figures~\ref{fig:csine}(b) and (c) show numerical results for the MSD
and the WTD of the climbing sine map randomized by using the scheme of
Fig.~1 in the main text. That is, with probability $1-p$ we choose the
parameter $c_1=0.8$ sampling the chaotic region while with probability
$p$ we draw from the attractive periodic orbit case at $c_2=0.2$. Both
figures display results for the critical probability $p_c\simeq
0.505702$, where the map's Lyapunov exponent is approximately
zero. One can clearly see excellent matching of the exponents $\sim
t^{1/2}$ for the MSD and $\sim t^{-3/2}$ for the WTD as predicted by
conventional CTRW theory, cf.\ Fig.~2 in the main text. Hence, the
basic mechanism that we propose to generate anomalous diffusion due to
randomization of deterministic dynamics is also robust for a generic
nonlinear map.

\end{appendix}


\begin{thebibliography}{54}
\bibitem[{iid()}]{iidens} \bibinfo{note}{We have computed the
      infinite invariant density starting from a normalised uniform
      initial distribution of $N=10^5$ points by iterating for
      $n=10^5$ time steps, which yields the right scaling in $x$
      \cite{KoBa09}, by not showing the cutoff at small $x$.}
\end{thebibliography}
\end{document}